\documentclass[aps,twocolumn,prx,superscriptaddress]{revtex4-2}
\usepackage{blindtext}
\usepackage{physics,siunitx}
\usepackage{standalone}
\usepackage{floatrow}
\usepackage{amsmath,amsfonts,amssymb}
\usepackage{graphicx}
\usepackage{amsthm}
\usepackage{multirow}

\usepackage{hyperref}
\usepackage{soul}
\usepackage{xcolor}

\begin{document}
	\title{Fast Flux Entangling Gate for Fluxonium Circuits}
	
	\author{Yinqi Chen}
	\email{chen447@wisc.edu}
    \affiliation{Department of Physics  and Wisconsin Quantum Institute, University of Wisconsin-Madison, Madison, Wisconsin 53706, USA}
	
	\author{Konstantin N. Nesterov}
	\altaffiliation{Present address: Bleximo Corp., Berkeley, CA 94710, USA}
    \affiliation{Department of Physics  and Wisconsin Quantum Institute, University of Wisconsin-Madison, Madison, Wisconsin 53706, USA}
    
    \author{Vladimir E. Manucharyan}
    \affiliation{Department of Physics, Joint Quantum Institute,
    and 
    Center for Nanophysics and
    \\ Advanced Materials,
    University of Maryland, College Park, Maryland 20742, USA}
    
    \author{Maxim G. Vavilov}
    \affiliation{Department of Physics and Wisconsin Quantum Institute, University of Wisconsin-Madison, Madison, Wisconsin 53706, USA}

    \date{\today}
    
\begin{abstract}

We analyze a high-fidelity two-qubit 
gate using fast flux pulses on superconducting fluxonium qubits. The gate is realized by temporarily detuning magnetic flux through fluxonium loop away from the half flux quantum sweet spot.
We simulate dynamics of two capacitively coupled fluxoniums during the flux pulses and optimize the pulse parameters to obtain a highly accurate $\sqrt{i\textsc{swap}}$-like entangling gate. 
We also evaluate the effect of the flux noise and qubit relaxation on the gate fidelity. Our results demonstrate that the gate error remains below $10^{-4}$ for currently achievable magnitude of the flux noise and qubit relaxation time.

\end{abstract}

    \maketitle
	\section{Introduction}

The fluxonium circuit is 
a promising  candidate for qubit implementation for a superconducting quantum processor~\cite{Manucharyan113}. In addition to having a strongly anharmonic spectrum, this qubit can exhibit millisecond-long coherence time because of the relatively low frequency of its main transition~\cite{Nguyenfluxonium,somoroff2021millisecond}, in comparison to transmons~\cite{Transmon}. Owing to the long coherence time, the single-qubit gate error has been recently demonstrated to be under $10^{-4}$~\cite{somoroff2021millisecond}.
Currently, the search for suitable two-qubit gates for fluxoniums is expanding in several directions that differ by the qubit control. The control can be realized by using microwave irradiation of the system~\cite{Nesterov_2018,Abdelhafez2020,Ficheux_2021, xiong2021arbitrary, Nesterov_2021, nesterov2022}, fast changes of the flux bias through the fluxonium superinductor loop~\cite{Zhang_2021,Bao2021}, or a tunable coupling scheme~\cite{Moskalenko2021,Moskalenko2022}.

The experimentally demonstrated two-qubit gate on fluxoniums was based on the microwave activation scheme~\cite{xiong2021arbitrary,Ficheux_2021}. In particular, a coupled two-qubit system was temporarily taken outside of the computational subspace, which results in a controlled-phase  operation~\cite{xiong2021arbitrary,Ficheux_2021}. 
Since noncomputational states generally have shorter lifetimes than the computational ones, these gate schemes are exposed to additional incoherent error channels. This problem can be avoided by utilizing schemes that keep the system entirely in the computational subspace during the gate operation.  Exploring ideas from transmon qubits~\cite{Transmon}, examples of microwave-activated gates with such a property include the cross-resonance gate~\cite{Chow2011, Sheldon2016b} and the two-photon b\textsc{swap} gate~\cite{Poletto2012}; both ideas have been analyzed theoretically for fluxonium qubits with promising predictions for possible error rate~\cite{Nesterov_2021,nesterov2022}.
Alternatively, the system can be kept in the computational subspace while two qubits are being entangled by bringing their frequencies in resonance using a rapid flux tuning. The disadvantage of this scheme is that the qubit is exposed to the flux-noise-induced decoherence while being off the sweet spot.  Nevertheless, flux-controlled 
single-qubit~\cite{Zhang_2021} and two-qubit~\cite{Bao2021} gates
have recently been demonstrated experimentally for fluxonium circuits.

In this paper we perform theoretical analysis of a fast-flux gate for two fluxoniums with a direct capacitive coupling and show that in the case of a unitary dynamics, such two-qubit gates  can  maintain a negligible gate error at a short gate time of less than 20 ns.  We focus on an $\sqrt{i\textsc{swap}}$-like gate, which mixes $\ket{01}$ and $\ket{10}$ states while also generally results in phase accumulation in $\ket{00}$ and $\ket{11}$ states.  This class of gates consists of  perfect entanglers, and $\sqrt{i\textsc{swap}}$  has been shown to provide powerful compilation capabilities rivaling those of $i\textsc{swap}$ and controlled NOT gates~\cite{huang2021ultrahigh}.

Frequency-tunable gates have been  analyzed and successfully implemented multiple times for weakly anharmonic superconducting qubits such as transmons~\cite{Strauch_2003, DiCarlo2009, Dewes_2012, Barends_2019,Wang_2019,Li_2019, Rol_2019, Foxen_2020, Sung_2021, Negirneac_2021}. There, gate schemes can be broadly divided into two categories depending on which resonance condition in two-qubit spectra is utilized. In the first category, weak anharmonicity is an asset  since it allows for an easy resonance condition between a computational and a higher energy noncomputational state such as $\ket{11}$ and $\ket{02}$, resulting in a controlled-phase gate implementation; see, e.g., Refs.~\cite{DiCarlo2009,  Negirneac_2021}. 
For fluxonium qubits, such a resonance is harder to achieve and is not desirable as mixing of computational and noncomputational states commonly results in increased leakage and decoherence. Instead, we analyze an operation from the second category of frequency-tunable gates, which are based  on a resonance between two computational levels~\cite{Dewes_2012, Barends_2019, Foxen_2020, Sung_2021}. This condition is achieved by flux biasing one of the fluxoniums  away from its sweet spot, which entangles the two qubits and can be tuned to realize an accurate $\sqrt{i\textsc{swap}}$-like gate. 
In addition to frequency-tunable implementations, $i\textsc{swap}$ and $\sqrt{i\textsc{swap}}$-like gates on superconducting quantum hardware have been also demonstrated via parametric activation~\cite{Abrams_2019, sete2021parametricresonance}. 

In comparison to weakly anharmonic qubits, where a similar gate scheme requires extra tuning of qubit frequencies or coupling constant to synchronize minima in the swap and leakage errors~\cite{Barends_2019}, this proposal for fluxonium qubits naturally suppresses coherent leakage to higher noncomputational levels because the system dynamics is adiabatic with respect to transitions to those levels.
This  resistance to leakage provides high fidelity for the gate, which is optimized  by adjusting parameters of the magnetic flux pulse, such as the pulse width, its raising and lowering times, and the pulse amplitude.
By choosing a proper flux detuning, the gate time can be shorter than $\SI{20}{\nano\second}$, with a coherent gate error below $10^{-6}$ and negligible leakage. Furthermore, the fluxonium qubits have the advantage to be insensitive to flux noise~\cite{Nguyenfluxonium}, thus making the gate noise resistant. To verify this statement and to account for other incoherent processes, we also analyze the effect of flux noise and  qubit relaxation on the gate fidelity.  We show that while these two effects result in significant reduction of the fidelity obtained from optimization of the qubit unitary dynamics, the gate error can still remain below $10^{-4}$ for currently reported values of the magnetic flux noise and relaxation times.

The paper is structured as follows. In Sec.~\ref{section:2qubit} we  describe the fluxonium Hamiltonian and present a simplified  analytic treatment based on two-level models. In Sec.~\ref{section:transition} we give a detailed account of how the entangling gate operation is realized, present numerical simulations of coherent dynamics of the gate as well as its coherent error. In Sec.~\ref{section:noise} we discuss reduction in gate fidelity due to flux noise and relaxation. We conclude in Sec.~\ref{section:conclusion}.

\section{Model of capacitively coupled fluxoniums}\label{section:2qubit}
\begin{figure}
\centering
\includegraphics[scale=0.15]{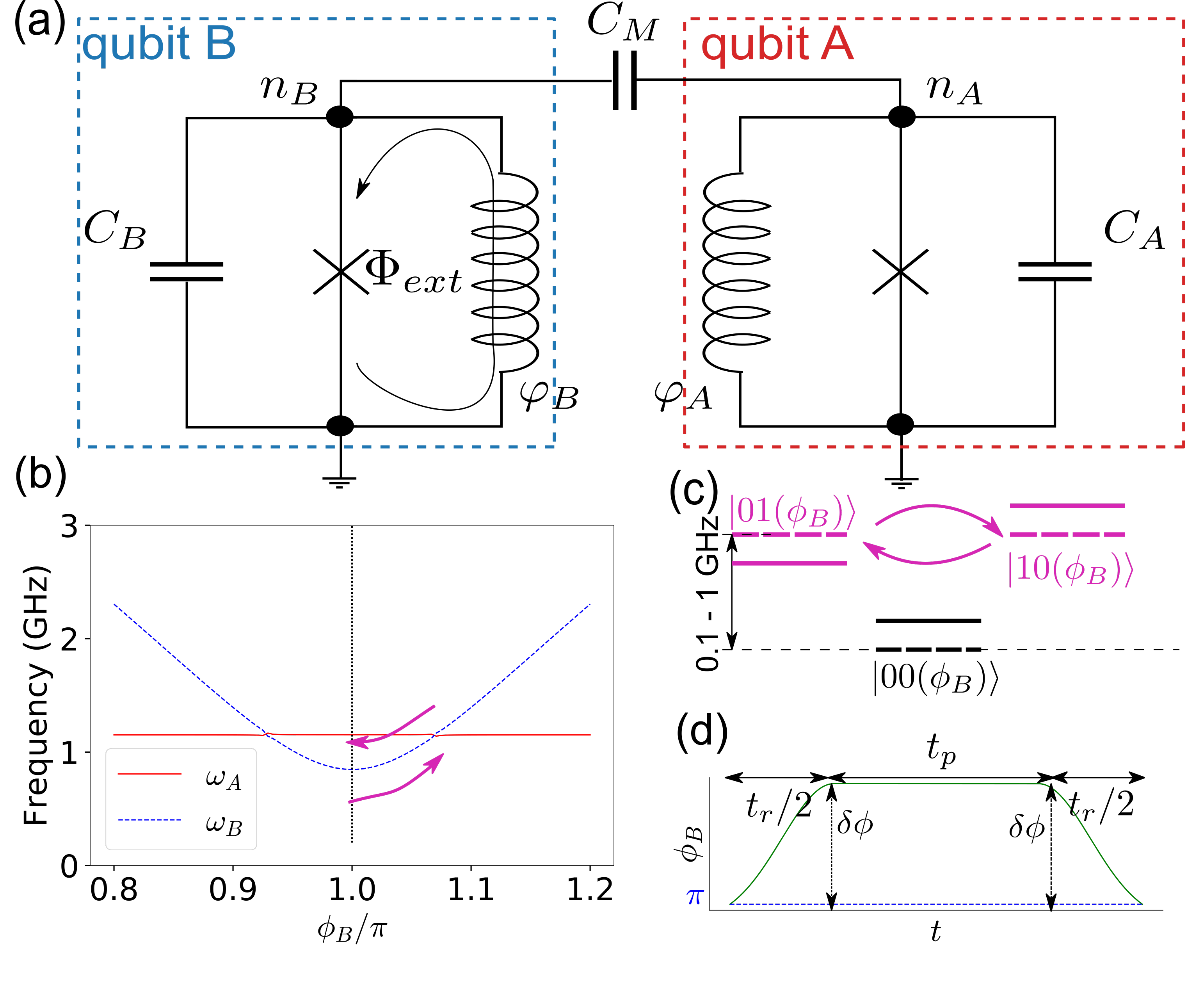}
\caption{Schematic of the fast flux entangling gate. (a) Circuit diagram of two capacitively coupled fluxonium qubits. (b) Single-qubit transition frequencies as a function of qubit $B$'s external flux. (c) Two-qubit energy levels at the flux sweet spot (solid lines) and at the flux  used for gate operation (dashed lines). (d) The flux-pulse shape represented by the flat-top Gaussian shape with switching on/off time $t_r/2$ and plateau duration $t_p$. 
}  \label{Fig:explanation}
\end{figure}

\subsection{Full Hamiltonian}
In this section, we introduce the model of interacting fluxonium qubits~\cite{Nesterov_2018, Nesterov_2021}.  Figure~\ref{Fig:explanation}(a) illustrates the circuit diagram of two capacitively coupled fluxoniums, labeled
as A and B. We
model this system by the Hamiltonian
\begin{equation}
\hat{H}=\hat{H}_A+\hat{H}_B+\hat{V}\,.
\label{eq:ham}
\end{equation}
Here
\begin{equation}
\hat{H}_{\alpha}=4E_{C,\alpha}\hat{n}_{\alpha}^2+\dfrac{1}{2}E_{L,\alpha}\hat{\varphi}_{\alpha}^2-E_{J,\alpha}\cos(\hat{\varphi}_{\alpha}-\phi_{\alpha})
\label{eq:qubits}
\end{equation}
describes individual fluxonium qubits $(\alpha=A,B)$\cite{Manucharyan113, fluxdependent}.  In this Hamiltonian, the canonical variables are flux $\hat{\varphi}_{\alpha}$ and charge (the number of Cooper pairs) $\hat{n}_{\alpha}$, which satisfy $[\hat{\varphi}_{\alpha},\hat{n}_{\alpha'}]=i\delta_{\alpha\alpha'}$. The kinetic term in  Eq.~(\ref{eq:qubits}) is determined by the charging energy $E_{C,\alpha}=e^2/2C_{\alpha}$, where $(-e)$ is the electron charge and $C_{\alpha}$ is the total capacitance of the circuit $\alpha$. The second term in Eq.~\eqref{eq:qubits} describes the superinductance or a long chain of Josephson junctions and depends on the inductive energy $E_{L,\alpha}=(\hbar/2e)^2/L_{\alpha}$, where $L_{\alpha}$ is the effective linear inductance of the chain. This superinductance is shunted by a small junction, which is characterized by a Josephson energy $E_{J,\alpha}$. The final term in Eq.~\eqref{eq:qubits} depends on $\phi_{\alpha}=(2e/\hbar)\Phi_{\alpha}$, where $\Phi_{\alpha}$ is the externally induced magnetic flux threading the loop formed by the small junction and superinductance. This parameter is tunable, and we use it to activate entangling gates.
We label single-qubit eigenstates as flux-dependent eigenstates of Hamiltonian \eqref{eq:qubits} as $\ket{k}_{\alpha,\phi_\alpha}$ for qubit $\alpha$ 
for the flux variable $\phi_{\alpha}$.  Here, 
$k=0,1,2,\dots$ is the excitation index of fluxonium labeling energy states $E^{(k)}_{\alpha,\phi_\alpha}$ in ascending order. 
The first two levels define the qubit transition frequencies $\hbar\omega_{\alpha, \phi_\alpha} = E^{(1)}_{\alpha,\phi_\alpha}-E^{(0)}_{\alpha,\phi_\alpha}$; we also write $\omega_\alpha = \omega_{\alpha, \pi}$.

We note that for a time-dependent external flux $\phi_\alpha(t)$, the formal circuit quantization procedure~\cite{You2019}
results in  an additional time-dependent term $(\hat{n}_\alpha/2\pi) d\phi_\alpha/dt$
in the
fluxonium Hamiltonian~\eqref{eq:qubits}. This term can be removed by choosing the 
degrees of freedom, when $\phi_\alpha$ is associated with the inductive term as $\frac 12 E_{L,\alpha} (\hat{\varphi}_{\alpha} + \phi_\alpha)^2$~\cite{You2019}. This addition is, however, negligible for the fluxonium parameters and flux variation speed studied in this paper with the small parameter controlling the approximation being $  h \partial_t \phi_\alpha/(16\pi E_C/h)$, see Appendix B in Ref.~\cite{Nguyen2022}. The term  can become important for faster changes in magnetic flux and the heavy fluxonium regime~\cite{Bryon2022}.

The interaction between the fluxoniums is described by
\begin{subequations}
\begin{equation}\label{eq:V}
\hat{V}=J_C\hat{n}_A\hat{n}_B,
\end{equation}
where the interaction constant $J_C$ is determined by the mutual capacitance $C_M$.
In the limit of a small mutual capacitance, $C_M\ll C_A,C_B$, the interaction strength is given by~\cite{Nesterov_2018,qec}
\begin{equation}
    J_C=4e^2C_M/(C_AC_B).
\end{equation}
\end{subequations}

Both qubits have sweet spots at $\phi_{\alpha}=\pi$, where the qubits are first-order insensitive to the flux noise and demonstrate long coherence times suitable for information storage and high-fidelity single-qubit operations~\cite{Nguyenfluxonium, somoroff2021millisecond}. 
Here we study a flux-tunable two-qubit gate that is activated by  moving the qubit with lower transition frequency away from the sweet spot for a short time. Specifically, we investigate in detail the scheme when qubit $A$ is kept at its sweet spot, so  $\phi_A=\pi$, while the magnetic flux drives qubit $B$ from the sweet spot towards the level crossing of both qubits; see Fig.~\ref{Fig:explanation}(b).
Since $\phi_A=\pi$ is fixed, we omit  index $B$ below for the flux variable $\phi_B=\phi$. 
We label interacting (dressed) two-qubit eigenstates of  Hamiltonian \eqref{eq:ham} as $\ket{kl}_{\phi}$, implying adiabatic connection to the noninteracting tensor-product
states $\ket{k}_A\ket{l}_{B,\phi}$.  The corresponding eigenenergies of the full Hamiltonian are denoted by $E_{\phi}^{(kl)}$. Three lowest two-qubit levels are shown schematically in Fig.~\ref{Fig:explanation}(c) for $\phi = \pi$ and for the value of $\phi$ corresponding to the level crossing.

\subsection{Hamiltonian in the computational subspace}

For any value of external flux $\phi$ used during the gate operation, the two-qubit computational
subspace $\{\ket{00}_{\phi},\ket{01}_{\phi},\ket{10}_{\phi},\ket{11}_{\phi}\}$ is separated from other states by a relatively large energy gap, suppressing the  leakage of the qubit state into higher noncomputational levels. 
 Higher noncomputational levels are generally important for quantitative analysis and are accounted for in numerical simulations in Secs.~\ref{section:transition} and \ref{section:noise}. To demonstrate the mechanism of the gate, we project the Hamiltonian into the noninteracting computational subspace fixed at the sweet spot $\phi=\pi$. In the  tensor-product basis $\{\ket{0}_A\ket{0}_{B,\pi},\ket{0}_A\ket{1}_{B,\pi},\ket{1}_A\ket{0}_{B,\pi},\ket{1}_A\ket{1}_{B,\pi}\}$,  we thus find that
\begin{equation}
\begin{aligned}
\label{eq:pauli}
\hat{H}_\phi & =-\dfrac{\hbar\omega_A}{2}\hat{\sigma}_{z,A}-\dfrac{\hbar\omega_{\phi}}{2}\hat{\sigma}_{z,B}+\dfrac{a_\phi}{2}\hat{\sigma}_{x,B}
+g\hat{\sigma}_{y,A}\hat{\sigma}_{y,B} \,.
\end{aligned}
\end{equation}
Here
\begin{subequations}
\begin{equation}
    \begin{aligned}
    \omega_\phi &= \omega_B -\dfrac{E_{J, B}(1+\cos\phi)}{\hbar}\\
    &\times\left[\bra{1}_{B, \pi} \cos\hat{\varphi}_B\ket{1}_{B, \pi}-\bra{0}_{B, \pi} \cos\hat{\varphi}_B\ket{0}_{B, \pi}\right]
    \end{aligned}
\end{equation}
and
\begin{equation}
    a_\phi =-2E_{J,B}\sin\phi\bra{0}_{B, \pi} \sin\hat{\varphi}_B \ket{1}_{B, \pi}
\end{equation}
describe diagonal and off-diagonal terms of qubit $B$'s Hamiltonian expressed in the $\phi=\pi$ basis
and 
\begin{equation}\label{eq:g_definition}
    g=\abs{J_C\bra{0}_{A}\hat{n}_{A}\ket{1}_{A}
\bra{0}_{B, \pi}\hat{n}_{B}\ket{1}_{B, \pi}}
\end{equation}
\end{subequations}
is the effective interaction strength in the computational subspace. {The representation of this Hamiltonian is similar to that of Ref.~\cite{ozguler2021excitation}, except in Ref.~\cite{ozguler2021excitation} the qubits have an inductive $XX$ coupling instead of a capacitive $YY$ coupling.

For vanishing interaction ($g=0$),
the energy gap $\hbar\Delta_\phi$ between crossing levels  
$\ket{01}_{\phi}$ and $\ket{10}_{\phi}$  is given by
\begin{equation} \label{eq:gap}
\Delta_{\phi}=\omega_A-
\sqrt{\omega_\phi^2+a^2_\phi/\hbar^2}\,.
\end{equation}
This gap gives the distance between the two lines in Fig.~\ref{Fig:explanation}(b) except that a realistic multi level fluxonium model was used for numerical results in the figure.
At the level crossing,  $\Delta_{\phi}$ vanishes and $a^2_\phi=\hbar^2|\omega_A^2-\omega_\phi^2|$.  

When  $g\neq 0$,  the interaction introduces an off-diagonal term in the reduced Hamiltonian in the two-dimensional subspace  $\mathcal{C} = \{\ket{0}_A\ket{1}_{B,\phi}, \ket{1}_A\ket{0}_{B,\phi}\}$, 
\begin{equation}
\hat{H}^{(\rm red)}(\phi)=-\frac{1}{2}\Delta_{\phi}\hat{\sigma}_z^{01, 10} + g\hat{\sigma}_x^{01, 10}\,.
\end{equation}
This Hamiltonian  describes  the Larmor precession in the subspace $\mathcal{C}$.
Our goal is then to quickly tune $\phi$ towards the level crossing $\Delta_\phi\approx 0$ and wait time $\sim h/g$ for states to evolve according  to a desirable two-qubit gate. This gate operation based on tuning energies of states $\ket{01}_\phi$ and $\ket{10}_\phi$ is illustrated schematically in Fig.~\ref{Fig:explanation}(c). 

Our gate scheme in the computational subspace is easier to understand in the limit when 
 changes of the magnetic flux are fast with respect to states $\ket{01}_\phi$ and $\ket{10}_\phi$, but  sufficiently smooth and adiabatic for states $\ket{00}_{\phi}$ and $\ket{11}_{\phi}$, so they only accumulate some phases after the gate operation but do not mix. These requirements imply 
 \begin{equation}\label{eq:gate_possibility_condition}
    g/\hbar \ll 1/t_r \ll \min(\omega_A, \omega_B)\,,
\end{equation}
 where $t_r$ is the total time spent on tuning magnetic flux during the operation; see Fig.~\ref{Fig:explanation}(d) for an example of a pulse shape. In numerical simulations discussed in Sec.~\ref{section:transition}, we find that high-fidelity optimized gates are possible even when limits \eqref{eq:gate_possibility_condition} are not strict, so $g/\hbar < 1/t_r < \min(\omega_A, \omega_B)$. In addition to this condition, in a realistic multi level fluxonium, the change of spectrum also has to be adiabatic with respect to transitions between computational and noncomputational levels such as between $\ket{11}_\phi$ and $\ket{12}_\phi$. This requirement is satisfied in the fluxonium because of its strong anharmonicity and, thus, a large spacing between computational and noncomputational levels.

\section{Unitary Dynamics }\label{section:transition}

By varying the external flux parameter $\phi=\phi_{B}$, we can modify the two-qubit spectrum. This way, as discussed in the previous section, see also Figs.~\ref{Fig:explanation}(b) and \ref{Fig:explanation}(c), the eigenenergies of states $\ket{10}_\phi$ and $\ket{01}_\phi$ can be quickly brought to the avoided-crossing point. While staying in the vicinity of this point, the state vector precesses in the $\ket{10}_\phi$ and $\ket{01}_\phi$ subspace and with proper timing, we can generate entangling gates.  In this section, we introduce $\sqrt{i\textsc{swap}}$-like gates and simulate  unitary dynamics of two coupled fluxonium qubits.

\subsection{$\sqrt{i\textsc{swap}}$-like gates}

We define an ideal target gate operation as the one where only mixing of states $\ket{01}_\pi$ and $\ket{10}_\pi$ occurs, while other transitions between basis states are not allowed.
Using single-qubit $Z$ rotations both before and after the operation, any unitary operator describing such a gate can be reduced to the form
\begin{equation}
\label{eq:Uid}
\hat{U}_{\rm ideal}(\theta,\zeta)=\begin{pmatrix}
e^{-i\zeta/2} & 0 & 0 & 0 \\
0 & \cos\frac\theta 2 & -i\sin\frac\theta 2 & 0 \\
0 & -i\sin\frac\theta 2 & \cos\frac\theta 2 & 0 \\
0 & 0 & 0 & e^{-i\zeta/2}
\end{pmatrix}\,.
\end{equation}
Here $\theta$ is the rotation angle for the subspace $\{\ket{01}_\pi, \ket{10}_\pi\}$, and the second angle, $\zeta$, describes the common effect of accumulated phases. In particular, $\zeta$ includes a contribution due to an effective $ZZ$ term~\cite{Zhao2020, Ku2020} in Hamiltonian~\eqref{eq:ham}. 
In Eq.~\eqref{eq:Uid} both angles $\theta$ and $\zeta$ are needed to parametrize gates up to single-qubit rotations.
In general, two gates \eqref{eq:Uid} with different pairs  $\theta$ and $\zeta$ cannot be reduced to one another by single-qubit rotations, i.e., they are from two different classes of local equivalence~\cite{Nesterov_2021}.
The family of local-equivalence classes given by Eq.~\eqref{eq:Uid} spans the family of excitation-preserving gates for fermionic simulations~\cite{Kivlichan2018, Foxen_2020}. Some prominent members of the family are \textsc{cz} [$U_{\rm ideal}(0, \pi)$], $\sqrt{i\textsc{swap}}$ [$U_{\rm ideal}(\pi/2, 0)$], $i$\textsc{swap} [$U_{\rm ideal}(\pi, 0)$], and \textsc{swap} [$U_{\rm ideal}(\pi, \pi)$] operations, which have different entangling properties. Using entangling power $\mathcal{P}$ to characterize these properties~\cite{Zanardi_2000, Ma2007}, we observe that \textsc{cz} and $i$\textsc{swap} have maximally possible entangling power of $\mathcal{P} = 2/9$, while  $\mathcal{P}=0$ for \textsc{swap} since it is not an entangling gate despite being a nonlocal operation. 

Here we consider the single-parameter subfamily of $\sqrt{i\textsc{swap}}$-like gates, which are parametrized by $U_{\rm ideal}(\pi/2, \zeta)$ with $\zeta$ varying between $0$ and $2\pi$. This group of gates includes $\sqrt{i\textsc{swap}}$ itself as well as $\sqrt{\textsc{swap}}$ [$U_{\rm ideal}(\pi/2, \pi/2)$] and $\sqrt{\textsc{swap}}^\dagger$ [$U_{\rm ideal}(\pi/2, 3\pi/2)$]. 
We observe that for $U_{\rm ideal}(\pi/2, \zeta)$ gates, $\mathcal{P}=1/6$ regardless of the phase $\zeta$, cf.~\cite{Nesterov_2021}. For the gates family \eqref{eq:Uid}, this robustness of $\mathcal{P}$ with respect to variations of $\zeta$ is a special property of $\theta=\pi/2$ and is not the case for other mixing angles. 

To find coherent gate fidelity for a realistic simulated unitary operator acting in a larger Hilbert space with  noncomputational levels and with possible leakage to those levels, we first project it into the computational subspace to obtain $\hat{U}_{\rm sim}$.  We calculate
\begin{equation}\label{eq:phase}
    \zeta = -\beta_{00} - \beta_{11} + \beta_{01} + \beta_{10}\,,
\end{equation}
where $\beta_{kl} = \mathrm{arg}\bra{kl} \hat{U}_{\rm sim}\ket{kl}$ is the diagonal-matrix-element phase of the simulated operator, and define the ideal target operator as $\hat{U}_{\rm ideal}(\pi/2, \zeta)$ according to Eq.~\eqref{eq:Uid}. Thus, an appropriate local-equivalence class is chosen each time a new computation is performed.
We then find $\hat{U}_{\rm sim}'$ by applying single-qubit $Z$ rotations to adjust phases of relevant matrix elements of $\hat{U}_{\rm sim}$ to make their structure be the same as in Eq.~\eqref{eq:Uid}. For example, we ensure that ${\rm arg}\bra{01} \hat{U}_{\rm sim}'\ket{01} = {\rm arg}\bra{10} \hat{U}_{\rm sim}'\ket{10} = 0$.
 Finally, we calculate coherent gate fidelity $F$ according to the standard expression~\cite{Pedersen2007},
\begin{equation}
F= \dfrac{\Tr(\hat{U}_{\rm sim}'^{\dagger}\hat{U}_{\rm sim}')+\abs{\Tr[\hat{U}_{\rm ideal}(\pi/2,\zeta)\hat{U}_{\rm sim}']}^2}{20}\,.
\end{equation}

\subsection{Optimization of the fast flux pulse}

\begin{table}
    \centering
\begin{tabular}{cccccc}
    \hline\hline
     \multirow{2}{*}{Qubit} & $E_{C, \alpha}/h$ & $E_{L, \alpha}/h$ & $E_{J, \alpha}/h$ & $\omega_\alpha/2\pi$ & $J_C/h$ \\
     & (GHz) & (GHz) & (GHz) & (GHz) & (GHz)\\
    \hline
    $A$ & 1.5 & 1.0 & 3.8 & 1.152 &  \multirow{2}{*}{0.3}\\
    $B$ & 0.9 & 1.0 & 3.0 & 0.848 & \\
    \hline\hline
\end{tabular}
\caption{Hardware parameters  used in numerical simulations.}
\label{table:para}
\end{table}

	\begin{figure}
	    \centering
	    \includegraphics[width=0.8\textwidth]{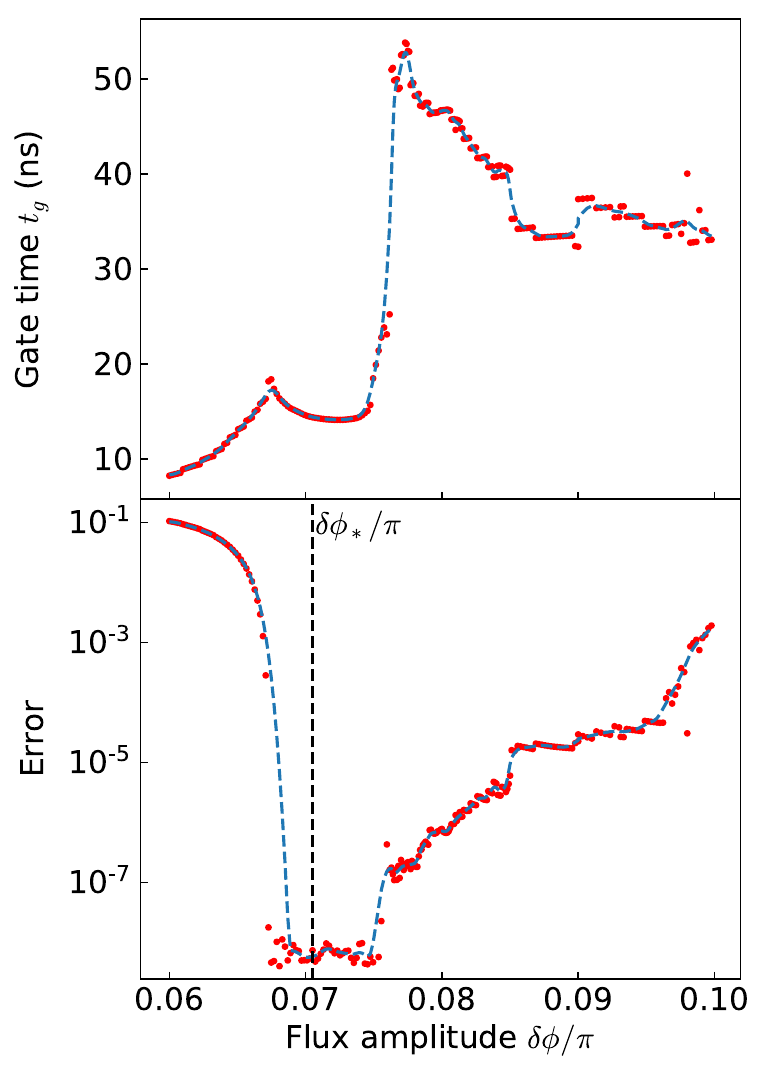}
	    \caption{Optimized total  gate duration (top panel) and coherent gate error (bottom panel) versus flux detuning at the plateau section of the flux pulse, see Eqs.~\eqref{eq:pulse} and \eqref{eq:pulse_plateau}. The red dots are the exact simulation data, and the dashed blue curve is a smooth fitting. Note an asymmetry around the avoided-level crossing detuning $\delta\phi_*\approx 0.0705\pi$.}
	    \label{Fig:inf}
	\end{figure}

 In this section, we present our analysis of the system dynamics described by the full Hamiltonian~\eqref{eq:ham} in response to time-dependent flux bias for fluxonium $B$, while keeping fluxonium $A$ at its sweet spot.
For numerical simulations, we use single-qubit parameters  as presented in  Table~\ref{table:para} and choose the coupling constant to be $J_C/h=\SI{0.3}{\giga\hertz}$. For these parameters, the fixed-$\phi_A$ avoided level crossing takes place at $\phi_{B}=\phi= \pi\pm \delta\phi_{*}$  with $\delta\phi_* \approx 0.0705\pi$. Detuning the external flux through fluxonium $B$ by a shift of $\delta\phi_{*}$, we can induce precession in the subspace of $\ket{01}_{\pi + \delta\phi_{*}}$ and $\ket{10}_{\pi + \delta\phi_{*}}$ states and generate  entanglement between the qubits, see Sec.~\ref{section:2qubit}. B. The avoided-level-crossing  energy splitting is $E_{10-01}/h=\SI{30}{\mega\hertz}$, meaning that the gate with $\theta = \pi/2$ in Eq.~\eqref{eq:Uid} can be faster than 20 ns. 

 To suppress  leakage to other states, we detune the flux of qubit B from the sweet spot using a smooth  Gaussian square pulse with ramp-up and -down times  $t_r/2$ each, plateau time $t_p$, and flux detuning at the plateau $\delta\phi$; see Fig.~\ref{Fig:explanation}(d). The pulse is then defined through
\begin{subequations}
\begin{equation}\label{eq:pulse}
\phi(t)=\pi+ C\delta\phi \left\{\exp\left[-\xi^2\dfrac{\bar{t}(\bar{t}-t_r)}{t_r^2}\right]-1\right\}
\end{equation}
when $0\leq t\leq t_r/2$ or $t_p+t_r/2\leq t \leq t_p + t_r$ and through
\begin{equation}\label{eq:pulse_plateau}
	    \phi(t)=\pi + \delta\phi
\end{equation}
\end{subequations}
when $t_r/2 \leq t \leq t_p + t_r/2$. Here
$C=[\exp(\xi^2/4)-1]^{-1}$ is the normalization factor, and the time variable $\bar{t}$ is defined as
	\begin{equation}
	    \bar{t}=
	    \begin{cases}
	    t & t\leq t_r/2\,,\\
	    t-t_p & t\geq t_r/2+t_p\,.
	    \end{cases}
	\end{equation}
We note that the actual  flux detuning $\delta\phi$ could deviate from the level crossing $\delta\phi_*$, while the gate fidelity remains high by adjusting plateau duration, which we shall demonstrate below.
		
For numerical simulations of gate dynamics, we use QuTiP package for PYTHON~\cite{Johansson2012, Johansson2013}. For each qubit, we first write Hamiltonian~\eqref{eq:qubits} in the basis of at least 30 harmonic-oscillator eigenstates of the same Hamiltonian but with $E_J = 0$. Then, we compute the eigenspectrum of the full single-qubit Hamiltonian~\eqref{eq:qubits} and choose $N=5$ lowest energy states at the sweet spot.  We verify that keeping more levels in the truncated single-qubit Hilbert space does not noticeably affect numerically simulated gate error. A larger $N$ is used for a detailed analysis of coherent leakage processes with the corresponding contribution to the total error being below $10^{-8}$, see Sec. \ref{sensitivity}. Using two sets of sweet-spot eigenstates for both qubits, we form the two-qubit tensor-product basis of $N\times N$ levels and the two-qubit Hamiltonian \eqref{eq:ham}. 
We perform all the computations in this tensor-product basis, which we refer to as the sweet-spot basis. To simulate the gate operation, we focus on the interacting computational subspace and find the final state for each eigenvector (or the density matrix) of that subspace. This way, we reconstruct the evolution matrix (or quantum process) of the gate, which is generally nonunitary if projected into the computational subspace. 

We first study the gate performance  as a function of flux detuning $\delta\phi$. For each value of $\delta\phi$, we optimize over pulse parameters $t_r,t_p,$ and $\xi^2$. We plot optimized gate duration $t_g= t_r + t_p$ and optimized infidelity versus $\delta\phi$ in Fig.~\ref{Fig:inf}. Below we identify and discuss three different regions defined by $\delta\phi$.

The first and main region of interest is the valley in the bottom panel of Fig.~\ref{Fig:inf}, defined by $0.067\pi\leq\delta\phi\leq0.075\pi$. In this valley, $\delta\phi$ is close to the avoided-level-crossing point $\delta\phi_*$, so the noninteracting gap~\eqref{eq:gap} is small and gate dynamics resembles that of an ideal gate discussed above. The plot shows that  the coherent gate error in the valley is below $10^{-7}$.
Such a high precision is sensitive to fluctuations of the pulse parameters. However, we demonstrate in Sec.~\ref{sensitivity} that the gate operated in this  regime provides a high fidelity above $99.99\%$ in a wide range of fluctuating parameters. This number is sufficient for quantum error correction~\cite{bravyi1998quantum, Fowler2012} and can help extending the depth of circuits executable on noisy intermediate-scale processors~\cite{Preskill2018quantumcomputingin}. In addition, the top panel of Fig.~\ref{Fig:inf} shows that the total gate duration in the valley is below 20 ns, implying that flux detuning pulses can deliver fast and high-fidelity $\sqrt{i\textsc{swap}}$-like gates.

For undershooting detuning,  $\delta\phi<0.067\pi$, the optimized gate fidelity drops fast below $0.9$. Since the avoided-level-crossing area is not reached for such detunings, amplitude $\lambda$ in expansion  $\ket{10}_\pi \propto \ket{10}_{\pi + \delta\phi} + \lambda\ket{01}_{\pi + \delta\phi}$ is not sufficiently large to produce a desired mixing of $\ket{01}_\pi$ and $\ket{10}_\pi$ by Larmor precession at $\phi = \pi + \delta\phi$.   In comparison,  when $\delta\phi = \delta\phi_*$, we find that $|\lambda|=1$, so an operation with any mixing angle $\theta$ is possible. The high-fidelity valley as in Fig.~\ref{Fig:inf} narrows for larger rotation angles $\theta$. Accurate gate operations with a smaller angle $\theta$ in Eq.~\eqref{eq:Uid} are possible in a wider region.

On the other hand, for overshooting detuning, $\delta\phi>\delta\phi_*$, the optimized gate fidelity can still exceed $0.999$. This high fidelity is possible because for a large detuning amplitude, the system  goes through  the avoided-level crossing and experiences the Landau-Zener transition between $\ket{01}_{\phi}$ and $\ket{10}_{\phi}$ states twice~\cite{Oliver2005, Berns2006}.  With proper phase accumulation between the transitions, a proper full evolution can be reduced to the desired form, Eq.~\eqref{eq:Uid}.

Note that the smallest coherent gate error found in our analysis is below $10^{-7}$; see Fig.~\ref{Fig:inf}(b). Leading digits in such small numbers may be affected by the terminating tolerance of our optimization protocol. This tolerance is chosen to balance the speed of computer simulations and the precision of our calculations. Here we focus on a general structure of how the flux detuning affects gate fidelity and on the demonstration that there is a wide range of parameters allowing a high fidelity. When decoherence effects are taken into consideration, the best gate error increases by several orders of magnitudes; see Sec.~\ref{section:noise} below.

\subsection{Gate dynamics in the high-fidelity region} 
\label{sensitivity}

Here we discuss gate dynamics in the high-fidelity valley of Fig.~\ref{Fig:inf} in more detail.
We first address the effects of timing and flux errors in pulse parameters; see Eqs.~\eqref{eq:pulse} and \eqref{eq:pulse_plateau}. To elaborate on this issue, we study gate fidelity as a function of $\delta\phi$ and $t_p$ with the ramp time and the Gaussian envelope parameter being fixed at $t_r=\SI{7.05}{\nano\second}$ and at $\xi^2=16.741$, respectively. 
These values of $t_r$ and $\xi^2$ are chosen as being close to the optimal values in the vicinity of $\delta\phi = 0.07\pi$ in the high-fidelity valley of Fig.~\ref{Fig:inf}. Plotting $1-F$ versus $\delta\phi$ and $t_p$ in Fig.~\ref{Fig:colormap}, we observe a triangle-shaped high-fidelity contour line with flux detuning $0.067\pi\leq\delta\phi\leq0.075\pi$. We find that with small variations in $\delta\phi$, coherent gate error can be kept below $10^{-6}$ by properly choosing plateau time along this high-fidelity contour. For successful error corrections, it is sufficient to have gate errors only below $10^{-4}$, which is also a realistic target goal given presently available best coherence times. For this error threshold, the allowed time interval error for plateau times for a given flux detuning is at least around $\SI{0.2}{\nano\second}$. The effects of flux errors are discussed in more detail in Sec.~\ref{section:noise}.

	\begin{figure}
	    \centering
	    \includegraphics[width=\textwidth]{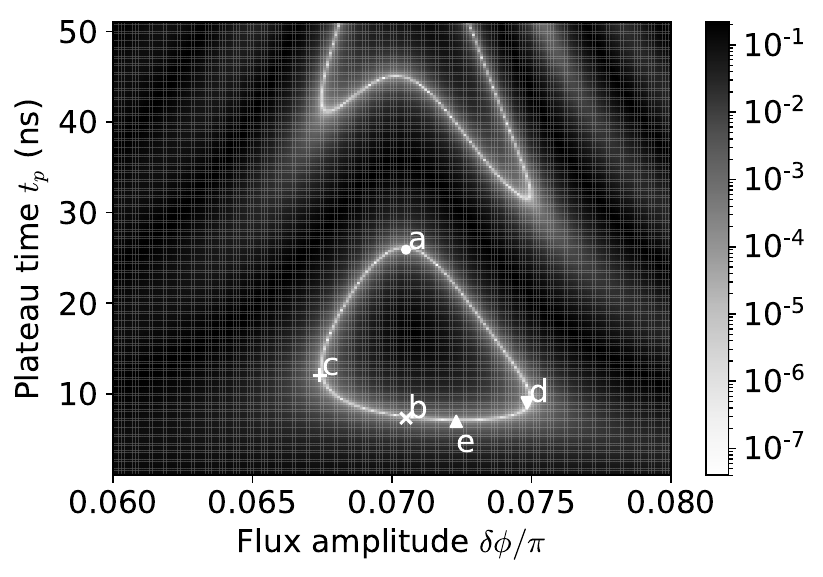}
	    \caption{Coherent gate error  versus flux-pulse parameters $\delta\phi$  and $t_p$ with other pulse parameters being fixed at $t_{r}=\SI{7.05}{\nano\second}$ and $\xi^2=16.741$. Points a and b are at the avoided-level crossing $\delta\phi=\delta\phi_*$. The evolution of state $\ket{01}_\pi$ at  parameters of points \textit{a}-\textit{d} is illustrated in time domain in Fig~\ref{Fig:bloch}. Parameters of points c and e, where point e is the bottom-most point of the high-fidelity contour, are used to plot Fig.~\ref{Fig:noise}.}
	    \label{Fig:colormap}
	\end{figure}
	
Along the high-fidelity contour, gate operations have similar entangling power as each of them is very close to $\hat{U}_{\rm ideal}(\pi/2, \zeta)$ for some $\zeta$. They, however, formally belong to different classes of local equivalence since they have different phases $\zeta$ calculated according to Eq.~\eqref{eq:phase}. Nevertheless, changes in $\zeta$ along the contour are less than $0.08\pi$ and are thus small with the values of $\zeta$ being primarily  determined by $t_p$ rather than $\delta\phi$.
To demonstrate how the dynamics differ along the contour, we pick four points to study evolution in time domain. Using the instantaneous basis
 $\{\ket{01}_{\phi(t)},\ket{10}_{\phi(t)}\}$, in Fig.~\ref{Fig:bloch}, we illustrate the evolution of eigenstate $\ket{01}_{\pi}$ of the full Hamiltonian~\eqref{eq:ham} by using the Bloch-sphere representation (left) and by plotting occupation probabilities (right). 

\begin{figure}
    \includegraphics[width=\textwidth]{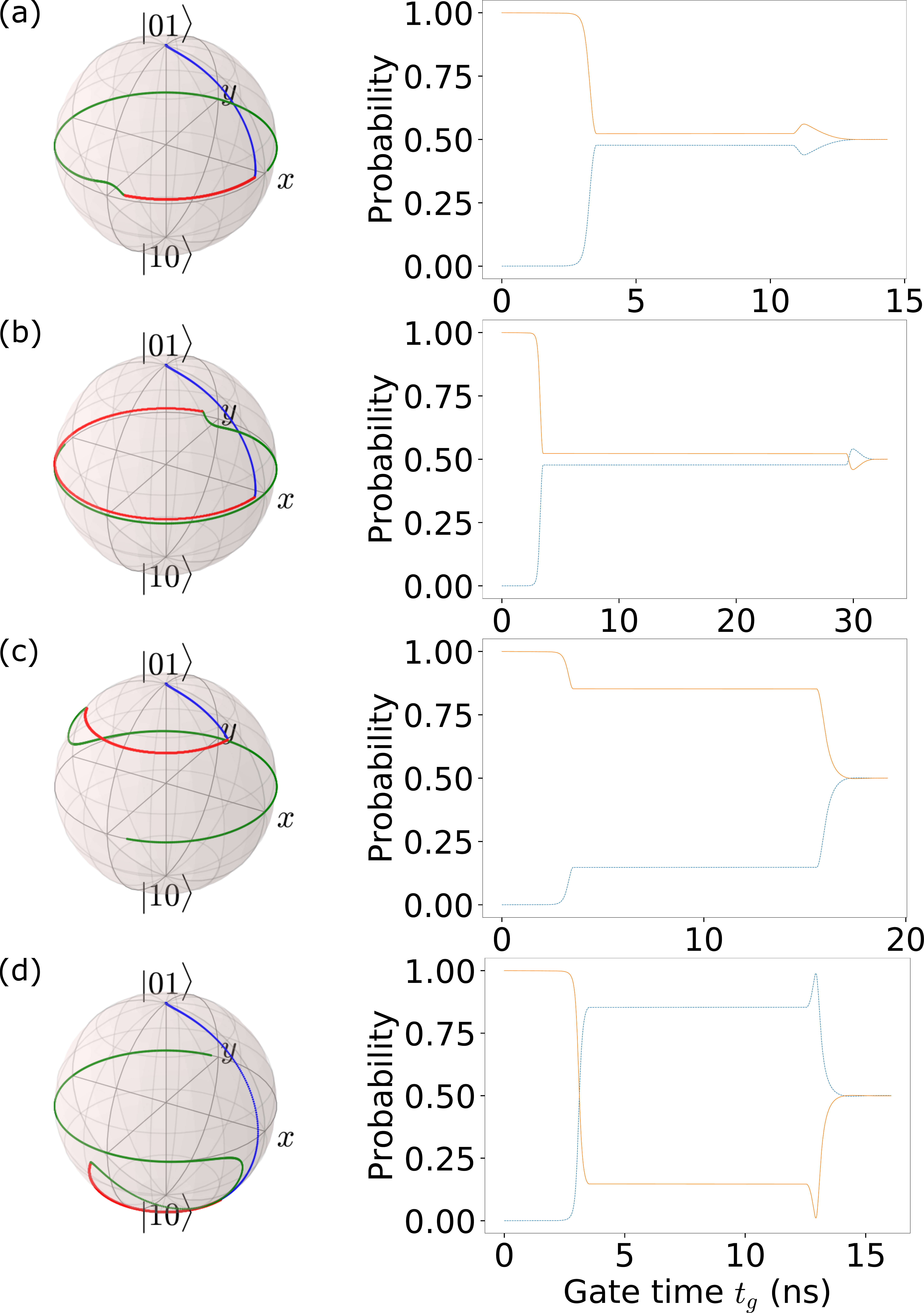}
    \caption{
    Unitary gate dynamics in the $\{\ket{10}_{\phi(t)}, \ket{01}_{\phi(t)}\}$ subspace for starting state $\ket{01}_\pi$. Left: Bloch-sphere trajectories for the ramp-on (blue), plateau (red), and ramp-off (green) parts of the flux pulse. Right: populations of states $\ket{01}_{\phi(t)}$ (solid orange lines) and $\ket{10}_{\phi(t)}$ (dashed blue lines). Panel labels correspond to pulse parameters of points a, b, c, and d in Fig. \ref{Fig:colormap}: (a) $t_{p}=\SI{25.85}{\nano\second},\delta\phi=0.0705\pi$; (b)  $t_{p}=\SI{7.30}{\nano\second},  \delta\phi=0.0705\pi$, (c)  $t_{p}=\SI{12.05}{\nano\second}, \delta\phi=0.0674\pi$; and (d)  $t_{p}=\SI{9.00}{\nano\second},\delta\phi=0.07482\pi$.
      Coherent gate fidelity exceeds $99.9999\%$ for each of the four points.
    }
    \label{Fig:bloch}
\end{figure}

Both Figs.~\ref{Fig:bloch}(a) and \ref{Fig:bloch}(b) demonstrate the gate operation for pulses with $\delta\phi = \delta\phi_*$, which bring the system precisely to the avoided-level crossing. Therefore, a fast ramp-up acting on state $\ket{01}_\pi$ results in an equal superposition of states $\ket{01}_\phi$ and $\ket{10}_\phi$, so the state is close to the 
Bloch-sphere equator during the plateau portion of the pulse.
Figures~\ref{Fig:bloch}(a) and \ref{Fig:bloch}(b) correspond to two significantly different plateau times $t_p$, which both generate high-fidelity gates, but with different relative phases of states $\ket{01}_\phi$ and $\ket{10}_\phi$. These phases accumulated on the equator of the Bloch sphere during the flat part of the pulse differ by $\pi$.  Figure \ref{Fig:bloch}(c) corresponds to the leftmost point of the high-fidelity contour in Fig.~\ref{Fig:colormap}, where $\delta\phi < \delta\phi_*$, so the state does not cross the equator of the Bloch sphere during the ramp-up portion of the pulse. In comparison, in Fig.~\ref{Fig:bloch}(d), $\delta\phi > \delta\phi_*$, so the state crosses the equator  during the ramp-up portion.
Both the probability plots and the Bloch spheres in Fig.~\ref{Fig:bloch} suggest that transitions between instantaneous states $\ket{01}_\phi$ and $\ket{10}_\phi$ happen only during the ramp-up and -down parts of the pulse. During the plateau, the amplitudes of $\ket{01}_\phi$ and $\ket{10}_\phi$ are constant, so the latitudes of the quantum states on the Bloch sphere remain constant as well. The function of the flat portion of the pulse is to wait for a state to precess in order to accumulate a proper phase difference between $\ket{01}_\phi$ and $\ket{10}_\phi$ in their superposition, so that the rotation happening during ramping down results in a proper final combination of $\ket{01}_\pi$ and $\ket{10}_\pi$. All the final states in Fig.~\ref{Fig:bloch} are located on the equator of the Bloch sphere, but have different phases, which can be absorbed into $\zeta$ of Eq.~\eqref{eq:Uid}. We however note that Bloch-sphere representations of quantum states in Fig.~\ref{Fig:bloch} do not describe additional $Z$ rotations that are used as a final step to reduce the operator to the standard form (\ref{eq:Uid}).
		
\subsection{Coherent leakage}  \label{section:leakage}

The main factors contributing to gate errors in superconducting qubits are leakage to noncomputational levels, flux noise, and the decoherence. One advantage of the fast flux-pulse gate is that the leakage can be suppressed in  fluxonium qubits, where the computational subspace is well separated from higher states, normally by several gigahertz, so flux pulses discussed in this paper do not drive state out of the computational subspace.

We estimate the leakage error of the gate by
\begin{equation}\label{eq:leakage_error}
    P_{\rm leak}=1-\dfrac{1}{4}\Tr{\hat{U}_{\rm sim}^{\dagger}\hat{U}_{\rm sim}},
\end{equation}
where $\hat{U}_{\rm sim}$ is the evolution operator projected to the  computational subspace. For parameters of Fig.~\ref{Fig:colormap}, we calculate $P_{\rm leak}$ versus $\delta\phi$ in the range $0.06\leq\delta\phi\leq 0.08$ at fixed $t_p=7.3$ ns and vs $t_p$ in the range $\SI{6}{\nano\second}\leq t_p\leq\SI{8}{\nano\second}$ at fixed $\delta\phi=0.0705$ and find that $P_{\rm leak} < 10^{-8}$ for all considered data points.
To check a detailed leakage profile, we increase the truncation cutoff to $N=10$ single-qubit levels for each qubit. We find that the main contribution to the leakage error~\eqref{eq:leakage_error} is coming from leakage into the higher states of the tunable qubit up to its fourth excited state, while the leakage probability to levels above $\ket{4}_B$ and to noncomputational levels of the fixed qubit is negligible in comparison to  $P_{\rm leak}$. Due to the low leakage probability, the main source of the gate error is induced by nonunitary processes in the computational subspace, which is discussed in Sec.~\ref{section:noise}. 
 
\subsection{Residual $ZZ$ Coupling}

A possible concern about multiqubit architectures with fixed coupling is the residual $ZZ$ coupling term, which leads to automatic entanglement and generally impairs both single- and two-qubit gate fidelities~\cite{Mckay2019}. This effective static interaction can be suppressed by using qubits with opposite-sign anharmonicities~\cite{Ku2020, Zhao2020}, in a multipath coupling scheme~\cite{Mundada2019, Kandala2021}, or by means of a  compensating $ZZ$ interaction that is induced dynamically by an off-resonant microwave drive~\cite{Wei2021, Mitchell2021, xiong2021arbitrary}.
In addition, $ZZ$ coupling can be steered off during the gate operation by using a special pulse design, see, e.g., Ref.~~\cite{Gambetta2020}. In this section, we investigate dependence of the static $ZZ$ coupling at the sweet spot and gate metrics on the coupling strength $J_C$.
 
 \begin{figure}
     \centering
     \includegraphics[width=0.8\textwidth]{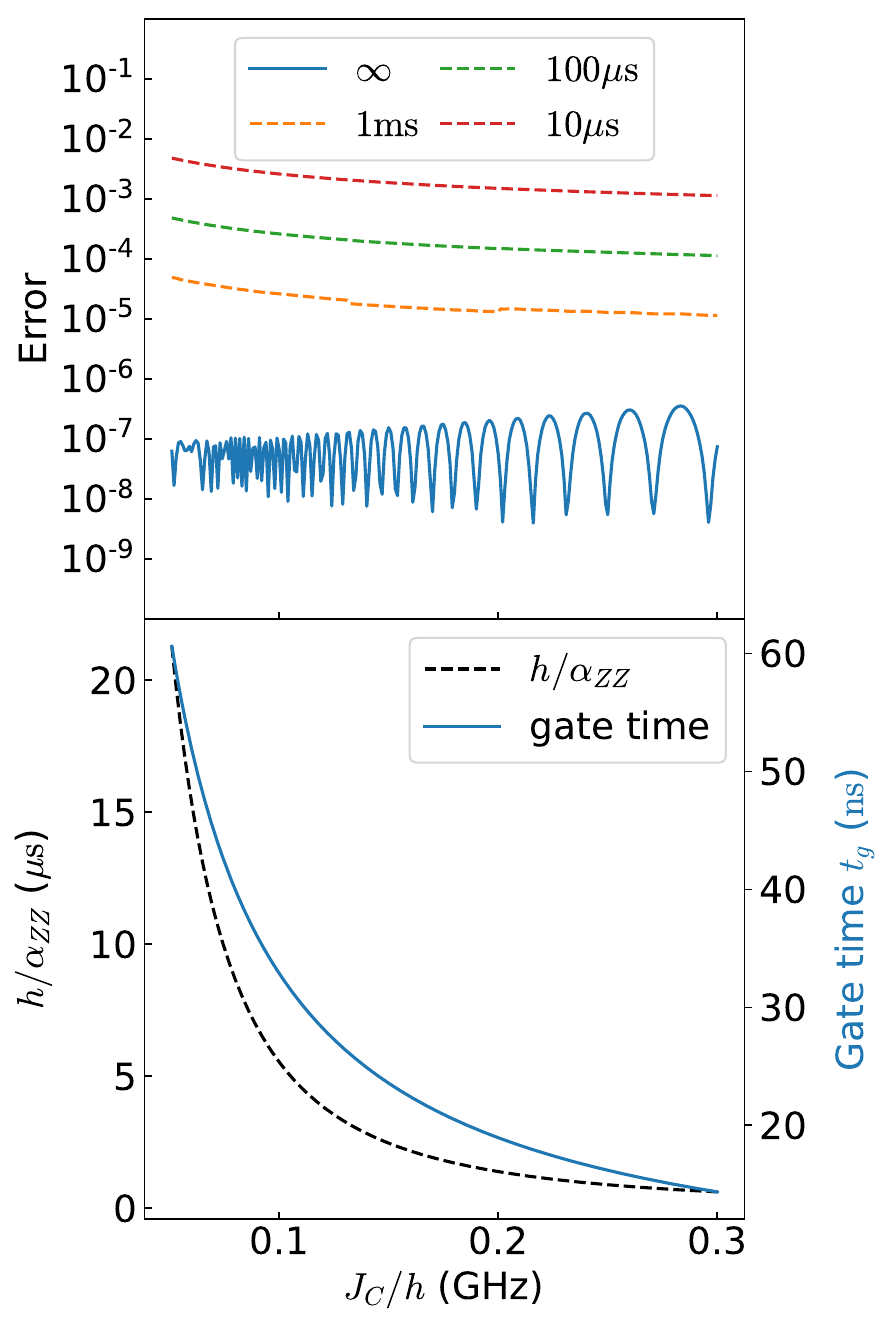}
     \caption{Top panel: coherent gate error (blue solid line, labeled as $\infty)$ and gate error in the presence of qubit relaxation with $T_1 = 1$ ms (orange), $T_1=100$ $\mu$s (green), and $T_1 = 10$ $\mu$s (red) vs $J_C/h$. The coherent error is optimized only over the plateau time $t_p$ with $t_r$ and $\xi^2$ taken as in Fig.~\ref{Fig:colormap} and $\delta\phi=0.0705\pi$. Bottom panel: optimized total gate time $t_g$ (blue solid line) and inverse $ZZ$ interaction rate $h/\alpha_{ZZ}$ (black dashed line) vs $J_C/h$.}
     \label{Fig:zz}
 \end{figure}
 
We note that according to our definition of the ideal gate operation~\eqref{eq:Uid}, coherent gate error is independent of $ZZ$ interaction as its effect can be incorporated into parameter $\zeta$. We illustrate this statement by numerically simulating this error versus $J_C/h$, see blue solid line in the top panel of Fig.~\ref{Fig:zz}.  For each value of $J_C/h$, we fix the flux detuning $\delta\phi=0.0705\pi$, choose $t_r$ and $\xi^2$ as in Fig.~\ref{Fig:colormap}, and optimize over a single parameter - the plateau time $t_p$.   The optimal total gate time $t_g$ is shown by blue solid lines in Fig.~\ref{Fig:zz}(b) along with the inverse $ZZ$ interaction rate $h/\alpha_{ZZ}$, which is shown by a black dashed line. 
Here we define the $ZZ$ rate according to
\begin{equation}
    \alpha_{ZZ}=E^{(00)}_\pi+E^{(11)}_\pi -E^{(10)}_\pi-E^{(01)}_\pi\,,
\end{equation}
where $E^{(ij)}_\pi$ stands for the energy of the sweet-spot eigenstate $\ket{ij}_{\pi}$.

Figure~\ref{Fig:zz} shows how coherent gate error can be kept consistently below $10^{-6}$ by optimizing over plateau time only, while the optimized $t_p$ varies inversely with $J_C$ as demonstrated in the bottom panel of Fig.~\ref{Fig:zz}. The latter observation is explained by noticing that the wait time scales as approximately $h/g$, where the effective coupling $g$ is proportional to $J_C$, see Eq.~\eqref{eq:g_definition} and Sec.~\ref{section:2qubit}. On the other hand, the $ZZ$ coupling is quadratic in $J_C$ from second-order perturbation theory. Thus, by reducing $J_C$  down to $\SI{0.05}{\giga\hertz}$, the value of $\alpha_{ZZ}$ can be made as small as 
$\SI{45}{\kilo\hertz}$ with $g$ still having a sufficiently large value of about $\SI{2}{\mega\hertz}$. At this value $J_C/h=0.05$ GHz, the coherent gate error is below $10^{-7}$ and the gate time is around $\SI{60}{\nano\second}$, see Fig.~\ref{Fig:zz}. Therefore, by choosing a smaller coupling constant $J_C$, it is possible to achieve a working high-fidelity entangling gate with low $ZZ$ coupling and a  short gate time below 100 ns. Incoherent gate error versus $J_C/h$ is discussed in Sec.~\ref{section:noise}.

\section{Nonunitary evolution}\label{section:noise}

Here we discuss effects of the environment on gate performance. Among these effects, two main sources of gate error are relaxation processes and low-frequency flux noise. 
We first focus on the gate error coming from relaxation.
	To account for these processes, we make an assumption that the decay channel is that of the single-qubit transitions at the sweet spot. We write the master equation on the density matrix $\hat{\rho}(t)$ as
	\begin{equation}
	    \begin{aligned}
	    \Dot{\hat{\rho}}(t)&=-\dfrac{i}{\hbar}\comm{\hat{H}(t)}{\hat{\rho}(t)}\\&+\sum_{\alpha=A,B}\left[2\hat{c}_{\alpha}\hat{\rho}(t)\hat{c}_{\alpha}^{\dagger}-\hat{\rho}(t)\hat{c}_{\alpha}^{\dagger}\hat{c}_{\alpha}-\hat{c}_{\alpha}^{\dagger}\hat{c}_{\alpha}\hat{\rho}(t)\right]\,,
	    \end{aligned}
	\end{equation}
	where $\hat{H}(t)$ is defined in Eq.~\eqref{eq:ham} and the collapse operators are given by
	\begin{equation}
	    \hat{c}_{A}=\dfrac{1}{\sqrt{T_{1,A}}}\sum_i\ket{0i}_{\pi}\bra{1i}_{\pi}\,, \hat{c}_{B}=\dfrac{1}{\sqrt{T_{1,B}}}\sum_i\ket{i0}_{\pi}\bra{i1}_{\pi}\,.
	\end{equation}
	Here we also assume that both qubits have identical relaxation times $T_{\alpha, 1} = T_1$.
    To find  fidelity for such a relaxation process, we first find the process fidelity $F_p=\Tr(\chi_{\mathrm{ideal}}\chi_{\mathrm{sim}})$~\cite{Chow_2009}, where $\chi_{\mathrm{ideal}}$ is the $\chi$ matrix for the ideal entangling gate and $\chi_{\mathrm{sim}}$ is the simulated $\chi$ matrix of the actual quantum process. We find the gate fidelity empirically to be $F_g=[4F_p+\Tr(\chi_{\mathrm{sim}})]/5$, which connects gate and process fidelity and accounts for leakage~\cite{Nielsen_2002}. Using optimal pulse parameters found for unitary evolution in Sec.~\ref{section:transition}, with the assumption that both qubits have $T_1$ of $\SI{100}{\micro\second}$, we find the best gate fidelity to be $F_g=99.99\%$. Thus, although incoherent processes substantially increase gate error up from $10^{-7}$,  we still obtain an error rate that is sufficient for quantum error corrections. The main reason for this high fidelity is that the total gate duration is almost $10^4$ times smaller than $T_1$; thus, the operation is completed before the decay processes become significant.
    
	\begin{figure}
        \includegraphics[width=\textwidth]{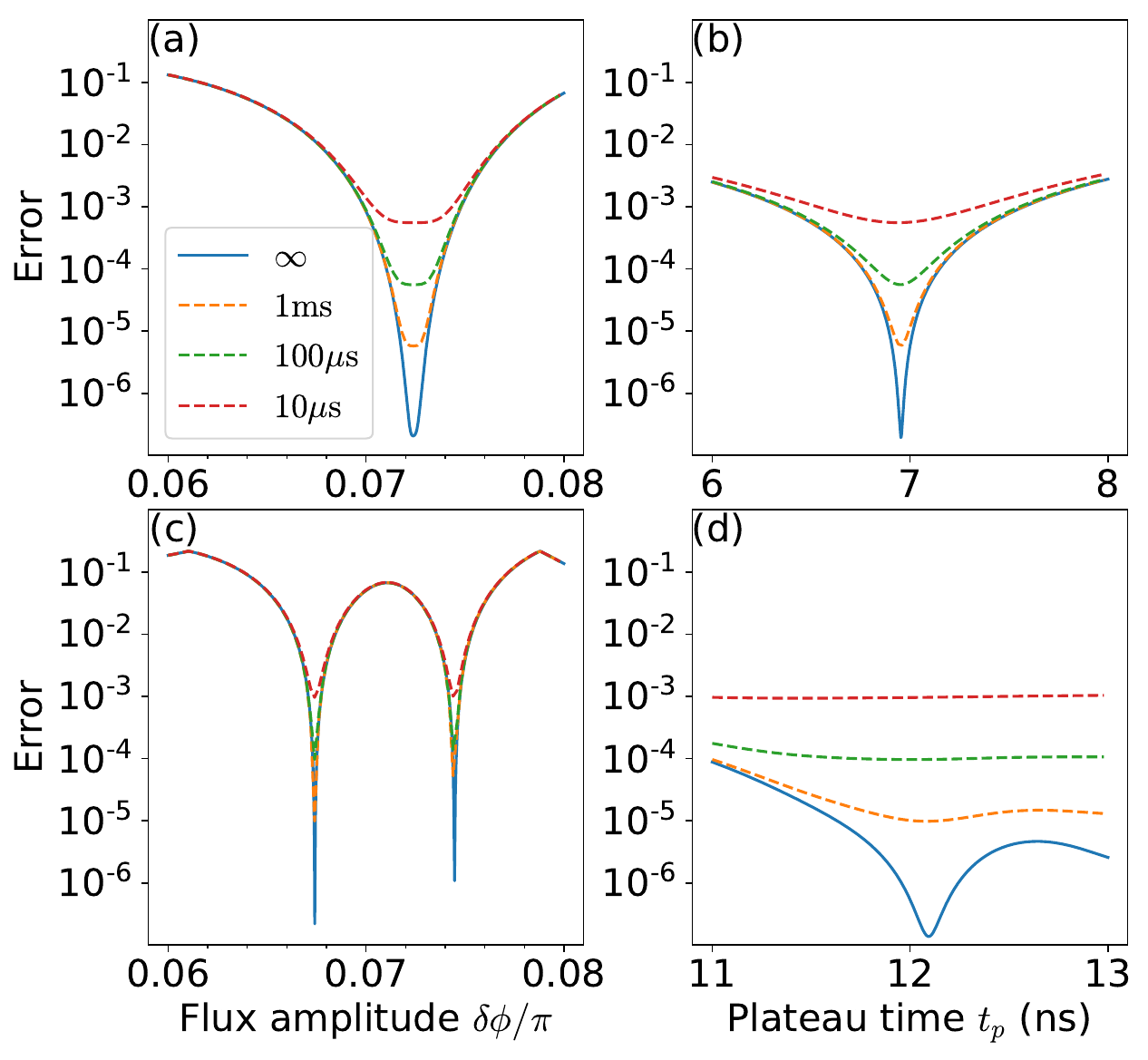}
        
	    \caption{Left column: the gate error vs $\delta\phi$, the flux detuning at the plateau, at a fixed  duration of the plateau $t_p=\SI{6.95}{\nano\second}$ (a) and $\SI{12.05}{\nano\second}$ (c). Right column: the gate error vs $t_p$ at a fixed $\delta\phi=0.0723\pi$ (b) and $0.0674\pi$ (d).
	     Shown are the results with no relaxation  (solid blue lines), with $T_1=\SI{1}{\milli\second}$, (orange dashed lines), $T_1=\SI{100}{\micro\second}$ (green dashed lines), and with $T_1=\SI{10}{\micro\second}$ (red dashed lines).  Fixed fluxes and times are chosen from the high-fidelity contour in Fig.~\ref{Fig:colormap} for points labeled as e (top row) and as c (bottom row). }
	    \label{Fig:noise}
	\end{figure}
	
In Fig.~\ref{Fig:zz}(a), we show how the total gate error is affected by the relaxation time $T_1$ as coupling strength $J_C$ varies. Since the gate time is approximately inversely proportional to $J_C$, see blue solid line in Fig.~\ref{Fig:zz}(b), a weaker coupling corresponds to a longer gate time, and the error accumulates according to  $1-e^{-t/T_1}\sim t/T_1$ law. Note that the oscillatory behavior displayed by the coherent error is absent  when $T_1$ relaxation is present since the decoherence takes lead over the imperfection in state rotation. The overall error is kept below $10^{-3}$ for $T_1=\SI{100}{\micro\second}$, and below $10^{-2}$ for $T_1=\SI{10}{\micro\second}$.
	
We now discuss gate error due to low-frequency flux noise. For fast flux-tunable gates discussed here, external magnetic flux $\phi_\alpha$ in Eq.~(\ref{eq:qubits}) is controlled by two sources and we express it as the sum  $\phi_\alpha = \phi_{\alpha, {\rm slow}} + \phi_{\alpha, {\rm fast}}$. The first contribution describes the source that enables relatively large values of $\phi_\alpha$, which cannot be changed fast, and also incorporates low-frequency flux noise. It is this source that is used to park qubits at the sweet spots, so, ideally, $\phi_{\alpha, {\rm slow}} = \pi$.
On the contrary, the second source allows for fast, but small changes in magnetic flux, and is used to quickly tune qubit $B$ away from its sweet spot, as described by the flux pulse in Eqs.~(\ref{eq:pulse}) and (\ref{eq:pulse_plateau}). At the sweet spot, where qubits are parked while idling and during single-qubit gates, we have $\phi_{\alpha, {\rm fast}} = 0$. Qubit transition frequencies are susceptible to low-frequency flux noise in $\phi_{\alpha, {\rm slow}}$, but only via second- and higher-orders sensitivities, resulting in a contribution to total qubit dephasing.  On the other hand, even if fast flux pulses $\phi_{B, {\rm fast}}(t)$ are stable between different gate operations, low-frequency noise in $\phi_{B, {\rm slow}}$ may result in starting values of $\phi_{B}$ at the beginning of the pulse being  different from $\pi$ and, therefore, may result in  $\phi_B$ at the plateau of the pulse exhibiting small fluctuations between different gate operations. The gate error due to this effect can be estimated
as the error due to miscalibration in the flux-pulse parameters. We present sensitivity to this type of control error in Fig.~\ref{Fig:colormap} and discuss it in more detail below.

In Fig.~\ref{Fig:noise} we combine the effects of both types of  errors discussed in this section. We show the dependence of gate error on the height of the flux pulse and duration of the plateau time for both unitary evolution (solid blue lines) and for two different relaxation times (dashed orange and dash dot green lines).
Top (bottom) row of Fig.~\ref{Fig:noise} shows the results for the horizontal and vertical line cuts at point e (point c) of the two-dimensional color plot of Fig.~\ref{Fig:colormap}. 
As expected, Fig.~\ref{Fig:noise} shows that gate error increases with decreasing $T_1$. However, even for a relatively short relaxation time of 10 $\mu$s, we observe gate errors that are only around $10^{-3}$ because of short gate durations. We also observe that the width of the valley around local minima of infidelity increases with decreasing $T_1$, which indicates that while the optimized error increases at shorter relaxation times,
the gate becomes less sensitive to control errors and errors due to flux noise. For relaxation times that are at least 100 $\mu$s, Fig.~\ref{Fig:noise}(a) suggests that the gate error remains below $10^{-4}$ in the presence of flux noise below $10^{-3}$ of the flux quantum $2e/h$. For other working points, e.g., Fig.~\ref{Fig:noise}(c), which corresponds to point c in Fig.~\ref{Fig:colormap}, the gate is more sensitive to flux noise, but flux noise $10^{-4}\times 2e/h$ is still compatible with the gate error below $10^{-4}$.

\section{Conclusion}\label{section:conclusion}
We present a way to build a fast $\sqrt{i\textsc{swap}}$-like gate  on fluxonium qubits using flux detuning with fidelity greater than $99.99\%$. The gate is turned on by tuning qubit frequency with an external magnetic flux to the avoided-level crossing point for energies of $\ket{01}$ and $\ket{10}$ states. We demonstrat gate operation via simulations with Gaussian flat-top flux pulses that enable correct mixing of states $\ket{01}$ and $\ket{10}$ and heavily suppressed probability of transitions between other states.
In particular, this fast flux-tunable gate for fluxonium qubits has the advantage of weak leakage out of the computational subspace without any additional steps. In comparison, flux-tunable gates with weakly anharmonic qubits may require more advanced techniques such as a net zero scheme to mitigate leakage~\cite{Rol_2019, Negirneac_2021}.

	We also investigate the effects of the flux noise. We find that the relaxation times of 100 $\mu$s and flux noise below $10^{-3}\times 2e/h$ are sufficient to keep the gate error below $10^{-4}$, while flux noise below $10^{-4}\times 2e/h$ provides more freedom in choosing parameters of the flux pulse. A more modest error threshold of $10^{-3}$ requires relaxation times of only 10 $\mu$s. 
	 We also make a detailed comparison of  gate dynamics at four different sets of flux parameters that result in high coherent fidelity. As displayed in Fig.~\ref{Fig:bloch}, their difference is determined by whether the state in instantaneous basis crosses the equator of the Bloch sphere or not.

In addition, we explore properties of two-qubit system for various values of coupling constant $J_C$. By decreasing $J_C$, the $ZZ$ coupling can be reduced to a small value below 50 kHz  with gate time increasing inversely proportionally, while the gate fidelity remains sufficiently high for error correction. With the effect of relaxation accounted for, the gate error with this small value of ZZ coupling is below $10^{-3}$ for $T_1=\SI{100}{\micro\second}$ and below $10^{-2}$ for $T_1=\SI{10}{\micro\second}$. We believe that this proposal provides a promising way for building a fast entangling gate with fluxonium qubits that is robust against flux noise, has practically zero leakage, and is not noticeably affected by qubit relaxation.
	
\begin{acknowledgements}

We acknowledge the support from ARO-LPS HiPS program (Grant No. W911NF-18-1-0146). V.E.M. and M.G.V acknowledge the Faculty Research Award from Google and fruitful conversations with the members of the Google Quantum AI team.  We use the QuTiP software package~\cite{Johansson2012, Johansson2013}  and perform computations using resources and assistance of the UW-Madison Center For High Throughput Computing (CHTC) in the Department of Computer Sciences. The CHTC is supported by UW-Madison, the Advanced Computing Initiative, the Wisconsin Alumni Research Foundation, the Wisconsin Institutes for Discovery, and the National Science Foundation.

\end{acknowledgements}

	\bibliography{reference}

\end{document}